\documentclass[useAMS,usenatbib]{mn2e}
\usepackage{graphicx}
\usepackage{epsfig}
\usepackage{subfigure}
\usepackage{color}
\usepackage{xcolor}
\usepackage{multirow}
\usepackage{threeparttable}
\usepackage{natbib}
\usepackage{amsmath,upgreek,amssymb}

\title[MACS\,2129: Precise strong lensing analysis]{Precise strong lensing mass profile of the CLASH galaxy cluster MACS\,2129}

\author[A. Monna et al.]
{\parbox{\textwidth}{ 
A. Monna$^{1,2}$\thanks{E-mail: amonna@usm.uni-muenchen.de},
S. Seitz$^{1,2}$,
I. Balestra$^{1,3}$,
P. Rosati$^{4}$,
C. Grillo$^{5,6}$,
A. Halkola$^{7}$,
S. H. Suyu$^{8,9,10}$,
D. Coe$^{11}$,
G. B. Caminha$^{4}$,
B. Frye$^{12}$,
A. Koekemoer$^{11}$,
A. Mercurio$^{13}$,
M. Nonino$^{3}$,
M. Postman$^{11}$,
A. Zitrin$^{14,15}$\thanks{Hubble fellow}
}\vspace{0.4cm}\\
\parbox{\textwidth}{$^{1}$University Observatory Munich, Scheinerstrasse 1, 81679 Munich, Germany\\
$^{2}$Max Planck Institute for Extraterrestrial Physics, Giessenbachstrasse, 85748 Garching, Germany\\
$^{3}$INAF, Osservatorio Astronomico di Trieste, via G. B. Tiepolo 11, I-34131, Trieste, Italy\\
$^{4}$Dipartimento di Fisica e Scienze della Terra, Universit\'a di Ferrara, via Saragat 1, I-44122 Ferrara, Italy\\
$^{5}$Dipartimento di Fisica, Universit\`a  degli Studi di Milano, via Celoria 16, I-20133 Milano, Italy\\
$^{6}$Dark Cosmology Centre, Niels Bohr Institute, University of Copenhagen, Juliane Maries Vej 30, 2100 Copenhagen, Denmark\\
$^{7}$Institute of Medical Engineering, University of L{\"u}beck, Ratzeburger Allee 160, D-23562 L{\"u}beck, Germany\\
$^{8}$Max-Planck-Institut f{\"u}r Astrophysik, Karl-Schwarzschild-Str. 1, 85741 Garching, Germany\\
$^{9}$Institute of Astronomy and Astrophysics, Academia Sinica, P.O. Box 23-141, Taipei 10617, Taiwan\\
$^{10}$Physik-Department, Technische Universit\"at M\"unchen, James-Franck-Stra\ss{}e~1, 85748 Garching, Germany\\
$^{11}$Space Telescope Science Institute, 3700 San Martin Drive, Baltimore, MD 21208, USA\\
$^{12}$Department of Astronomy/Steward Observatory, University of Arizona, 933 North Cherry Avenue, Tucson, AZ 85721,
USA\\
$^{13}$INAF, Osservatorio Astronomico di Capodimonte, Via Moiariello 16, I-80131 Napoli, Italy\\
$^{14}$Cahill  Center  for  Astronomy  and  Astrophysics,  California Institute of Technology, MC 249-17, Pasadena, CA 91125, USA\\
$^{15}$Physics Department, Ben-Gurion University of the Negev, P.O. Box 653, Be\'er-Sheva 84105, Israel\\
}}
\begin{document}
\label{firstpage}
\pagerange{\pageref{firstpage}--\pageref{lastpage}}
\maketitle

\begin{abstract}
We present a detailed strong lensing (SL) mass reconstruction of the core of the galaxy cluster MACSJ\,2129.4-0741 ($\rm z_{cl}=0.589$) obtained by combining high-resolution HST photometry from the CLASH survey with new spectroscopic observations from the CLASH-VLT (Very Large Telescope) survey. A background bright red passive galaxy at $\rm z_{sp}=1.36$, sextuply lensed in the  cluster core, has four radial lensed images located over the three central cluster members.  Further 19 background lensed galaxies are spectroscopically confirmed by our VLT survey, including 3 additional multiple systems.  A total of 31 multiple images are used in the lensing analysis.  This allows us to trace with high precision the total mass profile of the cluster in its very inner region ($\rm R<100$\,kpc). Our final lensing mass model reproduces the multiple images systems identified in the cluster core  with high accuracy of $0.4\arcsec$. This translates to an high precision mass reconstruction of MACS\,2129, which is constrained at a level of $2\%$. 
The cluster has Einstein parameter $\Theta_E=(29\pm4)$\arcsec\, and a projected  total mass of $\rm M_{tot}(<\Theta_E)=(1.35\pm0.03)\times 10^{14}M_{\odot}$ within such radius. Together with the cluster mass profile, we provide here also the complete spectroscopic dataset for the cluster members and lensed images measured with VLT/VIMOS within the CLASH-VLT survey.
\end{abstract}




\section{Introduction}
Clusters of galaxies are the largest gravitationally bounded structures in the Universe and play a fundamental role 
in testing cosmological models, investigating the formation and growth of structures of the Universe and properties of dark matter (DM). 
The combination of different independent techniques (stellar kinematics in the Brightest Cluster Galaxy, X-ray, weak and strong lensing, stellar kinematics, etc.)  allows to robustly constrain the DM  density profile of galaxy clusters  from the inner region up to large radial distances ($\rm \sim 5\, Mpc$) \citep[see][]{Sand2002,Sand2004,Newman2009,Newman2011,Umetsu2012,Umetsu2016,Biviano2013,Balestra2015}.
Gravitational lensing offers a unique technique to investigate the mass distribution of galaxy clusters, since it is generated by the total  mass of the lens, i.e. both the baryonic and dark matter components. In the strong lensing regime, when giant arcs and multiple images of background sources are generated, the mass distribution of the inner region of galaxy clusters can be reconstructed in great detail \citep[e.g. see][]{Eichner2013, Monna2015, Grillo2014}. 
In this context the Cluster Lensing And Supernovae survey with Hubble (CLASH, \citealt{Postman2012a}) and the CLASH-VLT survey, provide high resolution photometry and spectroscopic dataset to measure cluster mass profiles by combining  strong  and weak lensing analyses (see \citealt{Umetsu2012}, \citealt{Coe2012}, \citealt{Medezinski2013}).  In particular, the total mass profile of inner regions ($\rm R < 100-200 kpc$) of galaxy clusters can be derived with extremely high precision of few percentage using deep high resolution HST photometry, as that collected by  the CLASH and the deeper Hubble Frontier Fields (HFF, P.I. J. Lotz) surveys \citep[e.g., see ][]{ Jauzac2015, Grillo2014}.
\\ 
In this paper we present the strong lensing reconstruction of the core of the galaxy cluster MACS J2129.4-0741 (\citealt{Ebeling2007}, hereafter MACS\,2129) using the high resolution photometry from CLASH and new spectroscopy from the CLASH-VLT survey. 
This is a massive galaxy cluster ($M_{2500}=4.7\pm1.7\times10^{14}h_{70}^{-1}M_{\odot}$, \citealt{Donahue2014}), selected within the CLASH survey for its lensing strength. 
A peculiar bright red galaxy is sextuply lensed in its core, and the deflection of this multiple system is strongly affected by the gravitational action of the central galaxies. Six further multiple images systems are identified in the cluster core. These allow us to put strong constraints on the mass distribution in the innermost region of the cluster. \citet{Zitrin2011a} present a first strong lensing model of the cluster, which is subsequently  refined using the CLASH photometry in \citealt{Zitrin2015}, reproducing the multiple images with rms of $\sim2\arcsec$. Here we perform a more detailed analysis of the cluster core and take advantage of the new CLASH-VLT spectroscopic data to trace the cluster mass distribution with higher accuracy.

\begin{figure*}
 \centering
 \includegraphics[width=18cm]{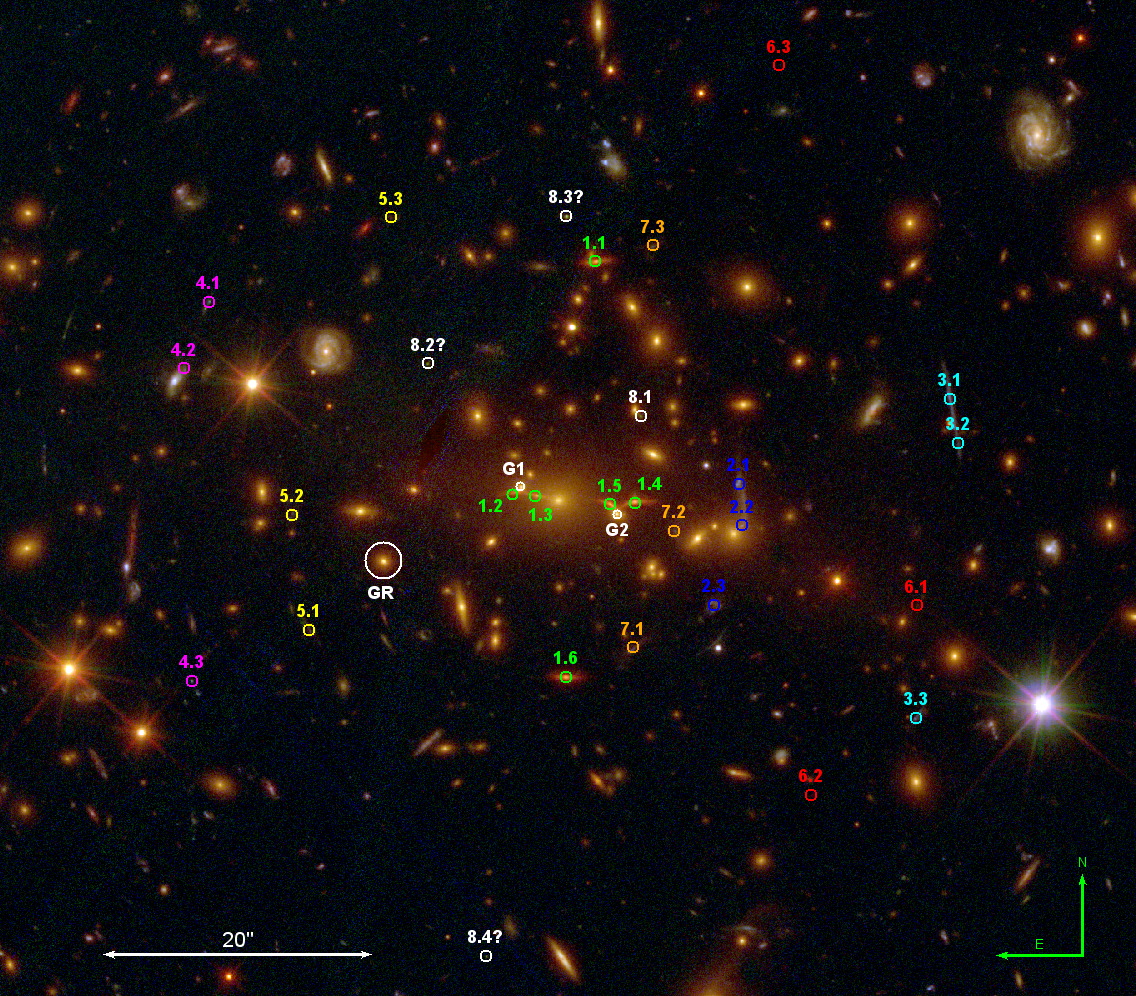}
 \caption{\small Color composite image of the core of MACS\,2129, generated using the CLASH HST dataset: Blue=F435W+F475W; Green=F555W+F606W+F625+F775W+F814W+F850LP; Red=F105W+F110W+F125W+F140W+F160W. 
 We label with different colors and numbers the seven multiple images systems used in the SL analysis. All systems have spectroscopic measurements except system 5 and 7. System 8 is a candidate quadruple lensed source. Image 8.1 was serendipitously discovered within our CLASH-VLT survey and has redshift $z_{sp}=4.4$. We selected three candidate counter images based on photometry, which are labelled as 8.2, 8.3 and 8.4.
 We also label the galaxy GR used as reference for the galaxy luminosity scaling relation and the two cluster members G1 and G2 which are individually optimized in our model.  }%
         \label{fig:fov}
 \end{figure*}

The paper is organized as follows.
In Section~\ref{sec:dataset} we present the photometric and spectroscopic datataset.
In Section~\ref{sec:lensing} we introduce the cluster mass components and the lensed systems used in the analysis. In Section~\ref{sec:point_model} we  describe the performed lensing analysis and in Section~\ref{sec:results}  we provide and discuss the results. 
Summary and conclusions are given in Section~\ref{sec:conclusions}. Finally in the Appendix we summarize the CLASH spectroscopy for the background sources and cluster members, and provide the 2D spectra and spectral energy distribution (SED) fit of the multiple images used in the lensing analysis. \\
Through the paper we assume a cosmological model with Hubble constant $H_0 = 70$ km s$^{-1}$ Mpc$^{-1}$,
and density parameters  $\Omega_{\rm m} = 0.3$ and $\Omega_{\Lambda}=0.7$. Magnitudes are given in the AB system. At the redshift of the cluster ($\rm z_{cl}=0.589$) 1\arcsec corresponds to 6.63 kpc.

\section{Photometric and Spectroscopic Dataset} 
\label{sec:dataset}
\subsection{ HST Photometry}
MACS\,2129 was observed between May and August 2011 as part of the CLASH survey with the HST Advanced Camera for Surveys (ACS) and the  HST Wide Field Camera 3 (WFC3) UVIS and IR cameras. These observations provide high resolution photometry in 16 filters covering the UV, optical and NIR range . In addition, photometry in the ACS/WFC/F555W filter is also available from  the public HST archive.
The photometric data set (available at http://archive.stsci.edu/prepds/clash, see \citet[][]{Postman2012}) is composed by HST mosaic drizzled 65mas/pixel images generated 
 with the \texttt{Mosaicdrizzle} pipeline \citep[see][]{Koekemoer2011}. These cover a field of view (FOV) of $\sim 3.5'\times3.5'$ in the ACS images and $\sim2'\times2'$ in the WFC3IR images.  
 In Tab.~\ref{tab:phot}  we list the complete filter list of our photometric dataset with the respective total exposure time and $5\sigma$ detection limit. \\
Using  these data,
 we generate a multi-band photometric catalog of  fluxes extracted within $0.6\arcsec$ (9 pixels) diameter aperture, using \texttt{SExtractor} 2.8.6 \citep{Bertin1996} in dual image mode. As detection image we use the weighted sum of all the WFC3IR images. 
 \begin{table}
\caption{Photometric Dataset summary for MACS\,2129: column (1) filters, column (2) total observation time in seconds, column (3) HST instrument, column (4) the $5\sigma$ magnitude depth within $0.6\arcsec $.}
\centering
\footnotesize
\begin{tabular}{|c|c|c|c|}
\hline
\hline
Filter&Total time (s)&Instrument&5$\sigma$ Depth\\
\hline
f225w &     6934 & WFC3/UVIS &   25.37\\
f275w &     7243 & WFC3/UVIS &   25.42 \\
f336w &     4580 & WFC3/UVIS &   25.78 \\
f390w &     4563 & WFC3/UVIS &   26.47 \\
f435w &     3728 & ACS/WFC   &   26.22  \\    
f475w &     4040 & ACS/WFC   &   26.64    \\  
f555w &     8880 & ACS/WFC&     26.97     \\
f606w &     3728 & ACS/WFC   &  26.86       \\
f625w &     3846 & ACS/WFC   & 26.37        \\
f775w &     4048 & ACS/WFC   & 26.24        \\
f814w &    13396 & ACS/WFC   & 27.14        \\
f850lp &    7808 & ACS/WFC   &  25.91      \\
f105w &  1006 & WFC3/IR   &    26.21  \\
f110w &  1409 & WFC3/IR   &    26.72  \\
f125w &  1409 & WFC3/IR   &    26.41    \\
f140w &  2312 & WFC3/IR   &    26.87    \\
f160w &  3620 & WFC3/IR   &    26.66    \\
\hline                  
\end{tabular}
\label{tab:phot}
\end{table}
\subsection{VLT Spectroscopy}
MACS\,2129 was observed with  the Visible Multi-Object Spectrograph \citep[VIMOS,][]{LeFevre2003} at ESO VLT as a part of the CLASH-VLT  ESO Program 186.A-0798   \citep[P.I. Rosati P., see][]{Rosati2014}. The cluster is partially obscured in the north-east side by a Galactic cirrus, thus only the central and South-western portion of the cluster were targeted by our VIMOS observations, covering a field of $16^\prime\times18^\prime$. Only two 4h-pointings were performed, one using the low resolution (LR) blue grism (covering the spectral range of 3700-6700\,\AA\, with a resolution of 180) and one using the medium resolution (MR) red  grism (which covers the 4800-10000\,\AA\, with resolution of 580). The spectra were reduced with the Vimos Interactive Pipeline Graphical Interface \citep[VIPGI][]{Scodeggio2005}  and redshifts were measured following the procedure described in \citealt{Balestra2015}. A quality flag (QF) is assigned to each spectroscopic redshift depending if it is ``secure" (QF=3), ``likely" (QF=2) , ``insecure" (QF=1) or if it is based on a strong emission line (QF=9). With a number of slits significantly reduced ($\sim1/10$) compared to other clusters observed in the CLASH-VLT campaign, we obtained a total of 281 redshifts, including 48 cluster members and 19 lensed sources. In the core of the cluster ($\rm R<1.5^\prime$), we have secured redshifts for 4 cluster members, 4 multiple images systems and 15 additional lensed background sources. In the appendix we provide the complete list of spectroscopic redshifts of cluster members and lensing features with their respective QF (see Tab.\,\ref{tab:galaxy_vlt} and \,\ref{tab:lensed_images}). \\

We also collected additional spectroscopic data available from literature \citep{Stern2010} which provide spectroscopic measurements of 6 further cluster members in the cluster core, observed with LRIS instrument on the Keck I telescope. 
Recently new spectroscopic measurements for cluster members and lensing features of MACS\,2129 has been released from the Grism Lens-Amplified Survey from Space \citep[GLASS, ][]{Treu2015}. However it provides no additional cluster member with robust $z_{sp}$ measurement to our galaxy sample.

\section{Strong Lensing Ingredients }
\label{sec:lensing}
We perform the strong lensing analysis in the core of MACS\,2129,  using the parametric mass modeling 
software \texttt{"Gravitational Lensing Efficient Explorer" (GLEE)}, developed by S. H. Suyu and A. Halkola \citep{Suyu2010,Suyu2012}. 
We adopt parametric mass profiles to describe the smooth large scale cluster dark halo (DH) and the cluster members.  The positions of the observed multiple images are used as constraints to estimate the mass profiles parameters. 
Through a simulated annealing minimization in the image plane we derive the best fitting model which reproduces the observed images. Then we perform  a Monte Carlo Markov Chain (MCMC) sampling to find the most probable parameters and uncertainties for the cluster mass components. 
 
\subsection{Cluster dark matter halo}
The large scale smooth DH of the cluster is described through a Pseudo Isothermal Elliptical Mass Distribution (PIEMD) profile \citep{Kassiola1993}, with projected surface mass density given by

 \begin{equation}
 \Sigma(R)= \frac{\sigma^2}{2 {\rm G}}(r_{\rm c}^2+R^2)^{-0.5}
 \end{equation}
where $\sigma$ is the halo velocity dispersion and $r\rm_c$ is the core radius. The projected radius R is given by  $R^2=x^2/(1+e)^2+y^2/(1-e)^2$ , where $e=(1-b/a)/(1+b/a)$ is the halo ellipticity, and $b/a$ is the halo axial ratio. For $(b/a\rightarrow1,r_c\rightarrow0)$, the asymptotic Einstein radius $\theta_E$  is given by
\begin{equation}
\theta_E=4\pi\left(\frac{\sigma}{c}\right)^{2}\frac{D_{ds}}{D_s}=\Theta_E\frac{D_{ds}}{D_s}
\label{eq:theta}
\end{equation}
where $D_{s}$ is the source distance and $D_{ds}$ is the distance between the lens and the source.  
In Eq.\,\ref{eq:theta} we introduce the Einstein parameter $\Theta_E$, which is the Einstein ring when the ratio $D_{ds} /D_{s}$   is 1. In our analysis the amplitude of the DH mass component is described through the Einstein parameter $\Theta_E$, which is optimized in the range $[5\arcsec,45\arcsec]$. The core radius $r_c$ is optimized within [0\arcsec,25\arcsec], while position angle PA and axis ratio vary within $[-90^{\circ},90^{\circ}]$ and [0,1], respectively. 
The position of the DH centre is optimized within $\pm3\arcsec$ around the central brightest cluster member (BCG). We adopt flat priors on all the DH parameters.

\subsection{Cluster members}
\label{sec:cmembers}

\begin{figure}
 \centering
 \includegraphics[width=9cm]{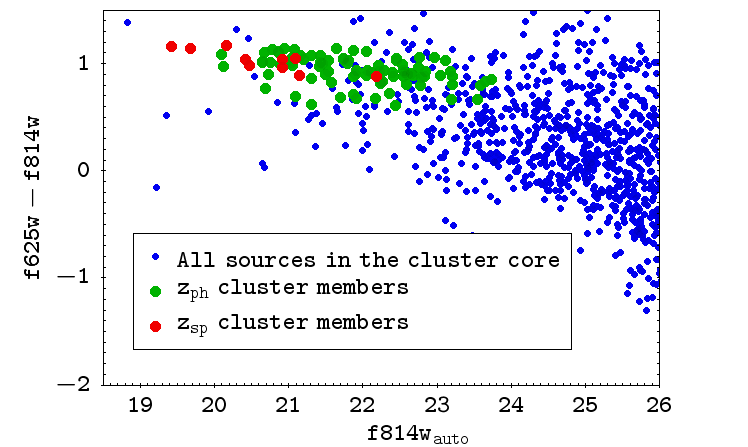}
 \caption{\small Color magnitude diagram for sources extracted in the  WFC3IR FOV of the cluster MACS\,2129 at $z=0.589$.
 We plot the color from the aperture magnitudes f625w and f814w, versus the \texttt{SExtractor} mag$\_$best in the f814w filter.
 Blue circles are all the sources extracted in the  WFC3IR FOV;
 Red circles are the spectroscopically confirmed cluster members ;
 Green circles are cluster member candidates with $z\rm_{ph}\in[0.52,0.65]$ and $\rm f814w\_mag\_best<24$.
 }  
 \label{fig:col_mag}
\end{figure}

 \begin{table}
\centering
\footnotesize
     \caption{List of the spectroscopically confirmed cluster members in the core of MACS\,2129: column (1) id, columns (2) and (3) RA and DEC in degrees, column (4) auto magnitude extracted with \texttt{SExtractor} in the F814w filter, column (5) spectroscopic redshift.}
\begin{threeparttable}
\begin{tabular}{|c|c|c|c|c|}
\hline
\hline
ID& RA&DEC &  $\rm F814W_{auto}$&z\\
\hline
  873 & 322.35712& $-$7.68909& $22.19\pm   0.01 $  &    0.583\tnote{a}  \\
  945 & 322.35495& $-$7.69178& $20.16\pm   0.01 $  &    0.596\tnote{a}  \\
  1298& 322.35095& $-$7.69715& $20.92\pm   0.01 $  &    0.593\tnote{a}  \\
  1747& 322.35324& $-$7.70795& $21.15\pm   0.01 $  &    0.590\tnote{a}  \\
  BCG & 322.3588 & $-$7.69105& $19.42\pm   0.01 $  &    0.589\tnote{b}  \\
  884 & 322.34265& $-$7.68953& $20.47\pm   0.01 $  &    0.590\tnote{b}  \\
  947 & 322.35574& $-$7.69188& $20.41\pm   0.01 $  &    0.586\tnote{b}  \\
  984 & 322.3631 & $-$7.69129& $20.92\pm   0.01 $  &    0.579\tnote{b}  \\
  1167& 322.35013& $-$7.69443& $21.10\pm   0.01 $  &    0.586\tnote{b}  \\
  1676& 322.3504 & $-$7.70633& $19.68\pm   0.01 $  &    0.596\tnote{b}  \\
  \hline                  
\end{tabular}

 \begin{tablenotes}
      \item[a] redshift measurement from CLASH-VLT survey
      \item[b] redshift measurement from \cite{Stern2010}
      \end{tablenotes}
\end{threeparttable}
      \label{tab:vlt_membrs}
\end{table}
In addition to the 10 spectroscopic members ( see Tab.\,\ref{tab:vlt_membrs}), we select further cluster members  in the core of the cluster ($\rm R<1.5^\prime$) using the CLASH photometry. These are  bright ($\rm f814w_{auto}<24$) galaxies which lie on the cluster red sequence ($\rm f625w-f814w\leq 1\pm0.3$), see Fig.\,\ref{fig:col_mag}. In addition we require that these candidate cluster members have photometric redshift  $\rm z_{ph}$ within $\rm z_{cl}\pm\Delta z$, with  $\rm \Delta z= 0.04(1+z_{cl})$. 
Photometric redshifts are computed based on the CLASH photometry  using the spectral energy distribution (SED) fitting code \texttt{LePhare}\footnote{http://www.cfht.hawaii.edu/$\sim$arnouts/lephare.html} \citep{Arnouts1999,Ilbert2006}. 
To estimate photometric redshifts we use galaxy SED templates from the COSMOS \citep{Ilbert2009}  template set.
To account for the interstellar medium extinction we apply the Calzetti extinction law \citep{Calzetti2000} to the starburst templates, and the SMC Prevot law \citep{Prevot1984} to the Sc and Sd templates.\\
We select 83 bright cluster member candidates in the cluster core based on the photometry. Two further galaxies are included in the sample after visual inspection, which have bluer color ($\rm f625w-f814w\sim0.6$) due to photometric contamination by nearby blue sources. We have a total of 85 candidate cluster members which add to the 10 spectroscopically confirmed ones.\\
These 95 galaxies are included in the cluster SL mass model and
are described with dual pseudo isothermal elliptical profiles (dPIE) \citep{Elisa2007}.  The projected surface mass density for this profile is  
\begin{equation}
 \Sigma(R)= \frac{\sigma^2}{2 {\rm G} R}\frac{r_{tr}^2}{r_{tr}^2-r_c^2} \left(\frac{1}{\sqrt{1+r_c^2/R^2}}-\frac{1}{\sqrt{1+ r_{tr}^2/R^2}}\right)
 \label{eq:bbs}
 \end{equation}
 where $R^2=x^2/(1+e)^2+y^2/(1-e)^2$ as for the PIEMD mass profile, $\rm r_c$ is the core radius,  $\rm r_{tr}$ is the so-called truncation radius which marks the region where the density slope changes from $\rho\propto r^{-2}$ to $\rho\propto r^{-4}$.
The total mass for this profile is given by
\begin{equation}
 M_{tot}=\frac{\pi\sigma^2}{G}\frac{r_{tr}^2}{r_{tr}+r_{c}}\,\,.
 \label{eq:m_bbs}
\end{equation}
We use vanishing core radii for the galaxies, thus, for each galaxy mass profile we have 2 free parameters, i.e. the velocity dispersion $ \sigma$ and the truncation radius $r_t$.\\
We adopt as reference galaxy (GR) one of the candidate cluster members, and then derive the velocity dispersion and size of all the other cluster members through the following luminosity scaling relations, according to the Faber-Jacksons and the Fundamental Plane relations:
\begin{equation}
\sigma=\sigma_{GR}\left(\frac{L}{L_{GR}}\right)^{\delta}  
\label{eq:F_J}
\end{equation}
\begin{equation}
r_{\rm tr}=r_{\rm tr,GR}\left(\frac{L}{L_{GR}}\right)^\alpha=r_{\rm tr,GR}\left(\frac{\sigma}{\sigma_{GR}}\right)^\frac{\alpha}{\delta}  
\label{eq:r_tr}
\end{equation}
Following our previous work, we adopt as exponent of the scaling relations $\delta=0.3$ and $\alpha=0.4$, assuming a constant mass to light ratio. See \cite{ Eichner2013, Monna2015} for a detailed derivation of these values.\\
The reference galaxy GR (Ra=322.3626, Dec=-7.69238 degrees, see Fig.\,\ref{fig:fov})  is a cluster member with magnitude $\rm F814w\_auto=21.3$ and axis ratio $\rm b/a=0.95$, which lies $\sim14\arcsec$ from the BCG. The velocity dispersion and truncation radius of GR are optimized with flat priors within the ranges of [100,300] km/s and [1,100] kpc, respectively.  The mass parameters $\sigma$ and $\rm r_{tr}$ for all the other cluster members are then derived through the previous scaling relation.\\
 The three central galaxies (the BCG and the two galaxies G1 and G2, lying on its sides, see Fig.\,\ref{fig:fov}) clearly affect the distortion of the radial multiple images of system 1, thus we individually optimized their parameters through the analysis. 
 The truncation radii for these galaxies vary within [1,100]kpc, the velocity dispersions are optimized in [100,300] km/s, except for the BCG for which we use a larger range of [100,410] km/s. In addition also their axis ratio $b/a$ and position angle PA are optimized, using the value extracted from the f814w image with gaussian priors.\\ For all the other galaxies $b/a$ and PA are fixed to the value extracted with Sextractor in the filter f814w.

\subsection{Lensed images}
\label{sec:multiple_images}
\begin{table*}
\caption{Summary of the multiple images identified in the core of  MACS\,2129: column (1) ID, column (2) and (3) RA and DEC in degrees, column (4) source redshift. Column (5) and (6) give the redshift of the lensed sources and the offset $\delta\theta$ between the observed and predicted multiple images as derived from our SL model, see Section \ref{sec:results}. All the systems have spectroscopic measurements, except for system 5 and 7, for which we provide the photometric redshift measurements from the online public CLASH archive, with their 68\% ranges.} 
\centering
\footnotesize
\begin{tabular}{|c|c|c|c|c|c|}
\hline
\hline
ID& RA& DEC& $\rm z_s$& $z_{sl}$& $\delta\theta (\arcsec)$\\
\hline
 1.1 &   322.35797 & -7.68588 & 1.36&1.36 &0.3 \\
 1.2 &   322.35965 & -7.69082 & "       &"&0.2 \\
 1.3 &   322.35925 & -7.69095 & "       &"&0.1 \\
 1.4 &   322.35712 & -7.69109 & "       &"&0.2 \\
 1.5 &   322.35764 & -7.69115 & "       &"&0.4 \\    
 1.6 &   322.35861 & -7.69489 & "       &"&0.6 \\  
 2.1 &   322.35483 & -7.6907  & 1.04&1.04 &0.2  \\
 2.2 &   322.35477 & -7.6916 & "       &" &0.1  \\
 2.3 &   322.35538 & -7.69332 & "       &"&0.4  \\
 3.1 &   322.35022 & -7.68886 & 2.24&2.24 &1.8  \\
 3.2 &   322.35004 & -7.68982 & "       &"&0.3 \\
 3.3 &   322.35095 & -7.69577 & "       &"&1.7  \\
 4.1a &   322.36642 & -7.68674 & 2.24&"   &0.4  \\
 4.1b &   322.36651 & -7.68689 & "      &"&0.3  \\
 4.2a &   322.36693 & -7.68831 & "      &"&0.4  \\
 4.2b &   322.36695 & -7.68820 & "      &"&0.2  \\
 4.3a &   322.36679 & -7.69497 & "      &"&0.3  \\ 
 4.3b &   322.36666 & -7.69525& "       &"&1.3  \\
5.1&	322.36422&	-7.69387&	1.8	[1.4,	2.0]&{$1.71\pm0.06$}&0.6\\
5.2&	322.36460&	-7.69137&	2.0	[1.8,	2.3]&"&0.3\\
5.3&	322.36243&	-7.68493&	1.8	[1.4,	2.2]&"&0.5\\
6.1&	322.35094&	-7.69333&	6.85	&6.85&1.4\\
6.2&	322.35324&	-7.69744&	"	&"&1.7\\
6.3&	322.35394&	-7.68164&	"	&"&1.9\\
7.1&	322.35714&	-7.69425&	1.4	[1.2,	1.5]&{$1.33\pm0.05$}&0.4\\
7.2&    322.35625&      -7.69172&      -                      &" &0.7\\
7.3&	322.35670&	-7.68554&	1.4	[1.1,	1.5]&"&0.5\\
8.1 &  322.35698   &-7.68924  &4.41  & 4.41& 0.5\\
8.2 &  322.36167   &-7.68808  &"  & "& 1.4\\
8.3 &  322.35860   &-7.68491  &"  & "& 1.2\\
8.4 &  322.36035   &-7.70094  &"  & "& 1.4\\
\hline
\end{tabular}
\label{tab:multiple_images}
\end{table*}

Eight systems of multiple images are clearly identified in the HST images of MACS\,2129, see Fig.\,\ref{fig:fov}. Seven of these were presented in \citet{Zitrin2015} (system 1 to 7), while system 8 was serendipitously discovered as a $\rm z_{sp}=4.4$ lensed source in the cluster core by our VLT observations. Within the CLASH-VLT survey, we targeted 21 lensed sources, including some belonging to the multiple images systems from \citet{Zitrin2015}, which were bright enough for VIMOS follow up. In Tab.\,\ref{tab:lensed_images} in the Appendix we list the redshift measurement for all the lensed features: 10 objects have secure  redshift measurement (QF=3), nine sources have likely redshift estimate (QF=2) and for two objects the $\rm z_{sp}$ are based on a strong emission line (QF=9). We obtain spectroscopic measurement for 5 systems of multiple images (systems 1 to 4 and system 8). In addition system 6 has been spectroscopically confirmed and analyzed in detail in \citet{Huang2016}. These spectroscopic measurements will be used as constraints in our strong lensing analysis.
Two further systems of multiple images are used in the SL modeling (system 5 and 7): these are identified by \citet{Zitrin2015} and have no spectroscopic measurement. The photometric redshifts of the multiple images associated to these systems are derived using the CLASH\footnote{available at http://archive.stsci.edu/prepds/clash/} photometry with the SED fitting code \texttt{Bayesian Photometric Redshift, BPZ} (see Tab.\,\ref{tab:lensed_images} and Fig. \ref{fig:dan_phot}). In the lensing analysis the mean $\rm z_{ph}$ for each system is used as starting value for the source redshift $\rm z_s$, and it is optimized with gaussian prior using as width the 68\% uncertainties on the photometric redshift estimates.  \\
In the following we give a short description of each lensed system. \\

\textbf{System 1.} A peculiar six times lensed red galaxy appears in the cluster core. Four multiple images are located on the sides of the BCG, and other two tangential images are $\sim16$\arcsec  northern and southern the BCG. Spectroscopic measurements of the image 1.1  and 1.5   place this galaxy at $\rm z_{sp}=1.36$. In Fig.\,\ref{fig:spectra1} the 1D and 2D spectra for images 1.1 and 1.5 are shown. The spectrum of this early type galaxy  shows a prominent 4000\,\AA\, break and no O[II] line in emission.\\%
\textbf{System 2.}
On the west side of the cluster, at a distance of $\sim14$\arcsec, a tangential arc and its counter image are found.
The arc has a spectroscopic redshift  of 1.04 with QF=2. The spectrum for this system is shown in Fig.\,\ref{fig:spectra1} \\
\textbf{System 3.}
At larger distance on the west side, there is a second long arc with respective counter image, which have likely spectroscopic redshift $\rm z_{sp}=2.24$ (QF=2). In Fig.\,\ref{fig:spectra2} we show the 1D and 2D spectra for images 3.1 and 3.3.\\
\textbf{System 4.}
A faint triply lensed source is identified on the eastern side of the cluster, at a distance of $\sim30$\arcsec. The three multiple images are clearly identified being composed by two bright knots. The image 4.3 was targeted with VIMOS, providing a likely source redshift of 2.24 (QF=2), see Fig.\,\ref{fig:spectra2}. \\
\textbf{System 5.}
This is a triply lensed system lying on the east side of the cluster core, at a distance of $\sim20\arcsec$ from the BCG. Photometric redshift estimate place this lensed galaxy at $\rm z_{ph}\sim1.5-2$.\\
\textbf{System 6.}
This system is composed by three optical dropouts with photometric redshift $\rm z_{ph}\sim6.5$. \citealt{Huang2016} have recently published the spectroscopic confirmation for this system using Keck/DEIMOS and HST/WFC3/G102 grism data. A clear Lyman$-\alpha$ emission line places this triply imaged system at $z_{sp}=6.85$. \\
\textbf{System 7.}
This system is another triply lensed galaxy which has photometric redshift estimation of $\rm z_{ph}\sim1.4$\\
 In Fig.\,\ref{fig:dan_phot} in the Appendix we show the optical and IR color cutout for the systems 5, 6 and 7, with their respective SED fit and  photometric redshift, as available from the public CLASH archive$^{2}$.\\
\textbf{System 8.} 
This is a candidate quadruply lensed source at $\rm z_{sp}=4.4$. Image 8.1 was serendipitously discovered in the vicinity of a cluster member targeted with our VIMOS observations. We obtained a spectroscopic redshift of $\rm z_{sp}=4.4$, based on a single emission line, identified as Lyman$-\alpha$ (see Fig.\,\ref{fig:spectra_z4}).  We photometrically identified  3 possible counter images of image 8.1 in the HST imaging which are supported by our lensing model, and are thus included in the model itself as constraints.\\ 

All the systems are summarized in Tab.\,\ref{tab:multiple_images}. In total we have 31 multiple images which provide constraints on our lensing model. 
We adopt uncertainties of $1\arcsec$ on the position of all the multiple images (except for system 1, see Sec.\,\ref{sec:point_model} for details), to take into account lens substructures and line of sight structures not considered in our model, which may introduce an offset of 0.5\arcsec-2.5\arcsec on multiple images prediction \citep{Host2012,Jullo2010}.   

\section{Strong Lensing Analysis}
\label{sec:point_model}
Given the mass components and the multiple images systems presented in the previous section, we perform the SL analysis of MACS\,2129 by minimizing the distance between the observed and predicted  multiple images. This is performed through a standard $\chi^2$ minimization, where the $\chi^2$ is defined as
\begin{equation}
\chi^2_{tot}=\sum\left(\frac{|\vec{\theta}^{pred}_i-\vec{\theta}^{obs}_{i}|}{\sigma^{pos}_i}\right)^2 \,,
\label{eq:chi_q}
\end{equation}
where ${\vec{\theta}^{pred}}_i$ is the predicted position of the $i-$th multiple image, $\vec{\theta}^{obs}_{i}$ is the observed multiple image position and $\sigma^{pos}_i$ is its uncertainty. 
We carry out the SL modeling through subsequent iterations, in which we include step by step the lensed systems in the model. \\
First, we perform the minimization analysis using as constraints only the two inner multiple systems with measured spectroscopic redshift (systems 1 and 2). Through this preliminary modeling we keep fixed the b/a and PA  of the BCG, G1 and G2. The position angle and axial ratio of these three galaxies are fixed to the values extracted with Sextractor in the filter f814w. The two lensed systems 1 and 2 provide 18 constraints through the positions of the respective multiple images. The free parameters are 18 in total: 6 for the DH, 2 for each of the individually optimized galaxy (GR, BCG, G1 and G2) and 4 for the x- and y- position of the lensed sources.  Once we reach a preliminary model with good prediction for the multiple images, 
 we add the outer spectroscopically confirmed systems 3, 4a and 4b. This increases the number of constraints to 36, while the free parameters are 24.
The inclusion of these outer multiple images affects the accuracy in reproducing the central system 1: the minimization is mainly driven by the tangential multiple images of systems 2, 3 and 4, which  place constraints on the large scale DH, to the detriment of the central radial images of system 1. In order to improve the mass model in the very inner region, then we reduce the positional error of system 1 to $\sigma^{pos}_{sys1}=0.5\arcsec$, which gives a higher weight to this  system in the $\chi^2_{tot}$ minimization. As a result of this approach, we reach preliminary  good prediction for all the spectroscopic confirmed systems, with $\rm \chi^2=16.9/12\,dof$ and median rms 0.5\arcsec. 
In the next step we include the multiple systems 5 to 8 with subsequent iterations and repeat the $\chi^2$ minimization. The source redshift for system 5 and 7 is unknown, thus we use their photometric redshift as starting value, and optimize them using a gaussian prior with width given by their respective $\rm z_{ph}$ uncertainties. At this stage we release the b/a and PA of BCG, G1 and G2 and  optimize within the range described in the previous section. In addition we also allow the core radii of these three galaxies to vary within [0,3.5] kpc. The total number of degrees of freedom is 27, having finally 62 constraints and 41 free parameters.
Besides the minimization analysis, we run MCMC chains to obtain the model which reproduces with best accuracy the observed multiple images and the uncertainties on our free parameters, i.e. the mass components parameters and the source redshift for systems 5 and 7.  

\section{Results and Discussion}
\label{sec:results}
\begin{figure}
 \centering
 \includegraphics[width=8.5cm]{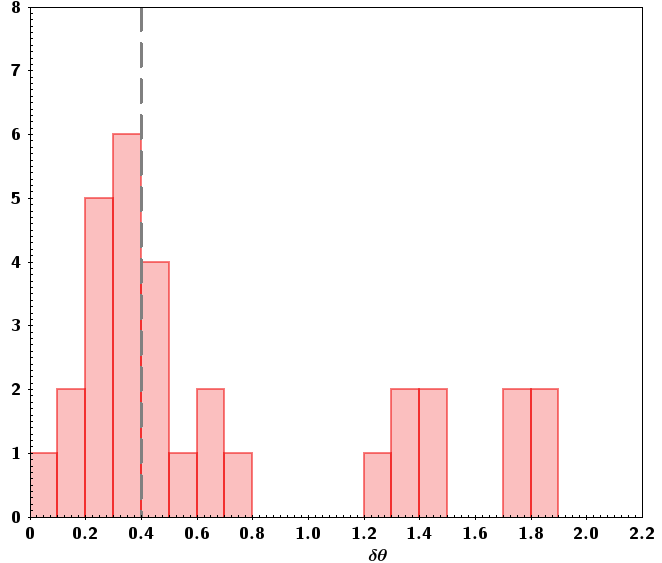}
 \caption{\small Histogram of the offsets $\delta\theta$ between the observed position of the multiple images and the predicted ones from our best SL model. Units are arcseconds. The vertical dashed line mark the median value of $\delta\theta$.}  
         \label{fig:rms}
 \end{figure}

\begin{figure}
 \centering
 \includegraphics[width=9cm]{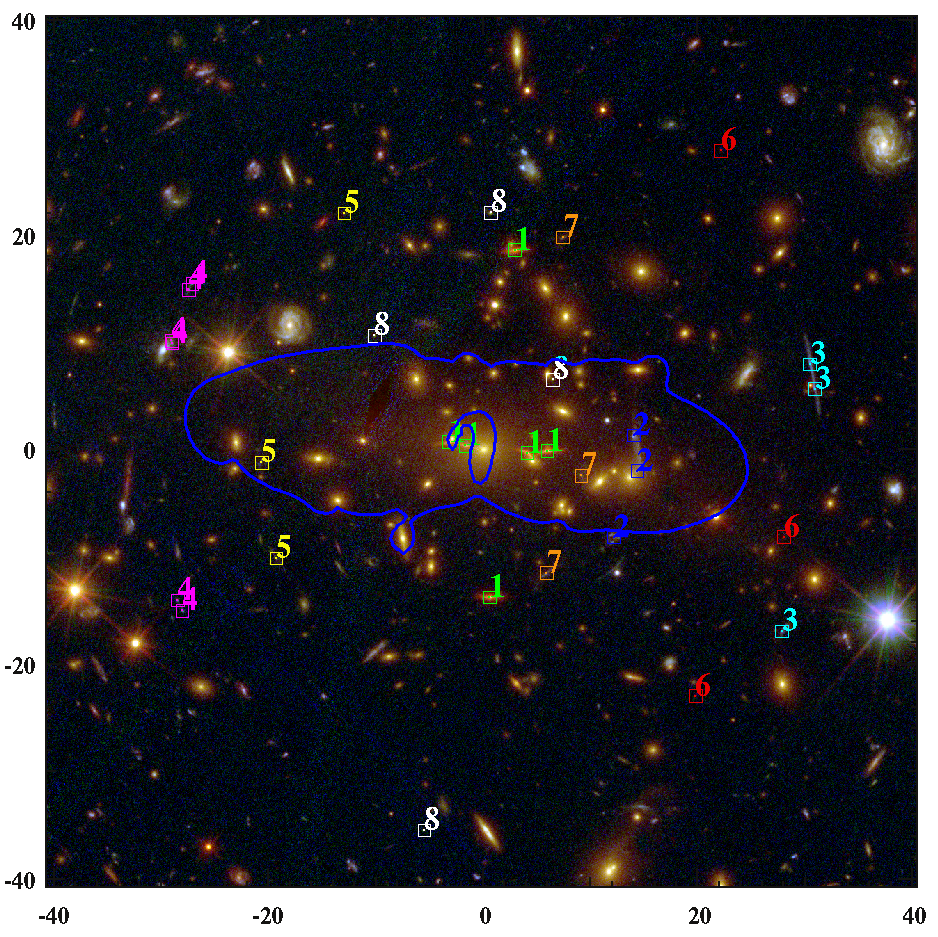}
 \caption{\small RGB color composite image of MACS\,2129 with overplotted the critical lines (in blue) for a source at the redshift $z_s=1.36$. The multiple images used to constrain the cluster mass profile are labeled with squares in different colors (see Tab.\ref{tab:lensed_images}). The axis are in arcseconds with respect to the BCG position }  
         \label{fig:cl_sys1}
 \end{figure}

After the minimization and MCMC analyses, our final best mass model of the cluster has $\chi^2_{tot}=29$ having 21 degrees of freedom. It reproduces the multiple images with a median accuracy of $0.4^{\prime\prime}$. 
In Fig.\ref{fig:rms} we show the histogram of the offsets $\delta\theta$ between the observed and predicted images, which are listed  in Tab.\ref{tab:multiple_images}.
In the inner region, inside the Einstein parameter of the cluster, the lensed systems 1, 2, 5 and 7 are reproduced with high accuracy, having median  $\delta\theta=0.4\arcsec$. Only the central system 8, which is a candidate quadruply lensed galaxy, has median $\delta\theta=1.3\arcsec$. The images 8.1, which is spectroscopically confirmed at redshift $\rm z_{sp}=4.4$, is reproduced with a small offset of 0.5\arcsec, whereas the three candidate counter images, photometrically selected, are predicted with accuracy of 1.3\arcsec. At larger radii ($\rm r\gtrsim30\arcsec$), we find that the two systems on the west of the cluster  (system 3 and 6) are reproduced with median $\delta\theta=1.7\arcsec$, while on the east side, System 4 is reproduced with high accuracy of  $0.4\arcsec$.
Based on the symmetry of the model (see Fig.\,\ref{fig:cl_sys1}), we can exclude that the particular shallowness of the DH profile is affecting the lensing prediction at large radii. Indeed, if this is the case, we would expect large $\delta\theta$ also for System 4. Thus we conclude  that  the larger offset $\delta\theta=1.7\arcsec$ for the prediction of system 3 and 6 is likely related to lens substructures or density inhomogeneities along the line of sight, which are not taken into account in our mass model. However  our photometric dataset does not reveal any apparent intervening structure in this region.\\
  
\begin{figure}
 \centering
 \includegraphics[width=8cm]{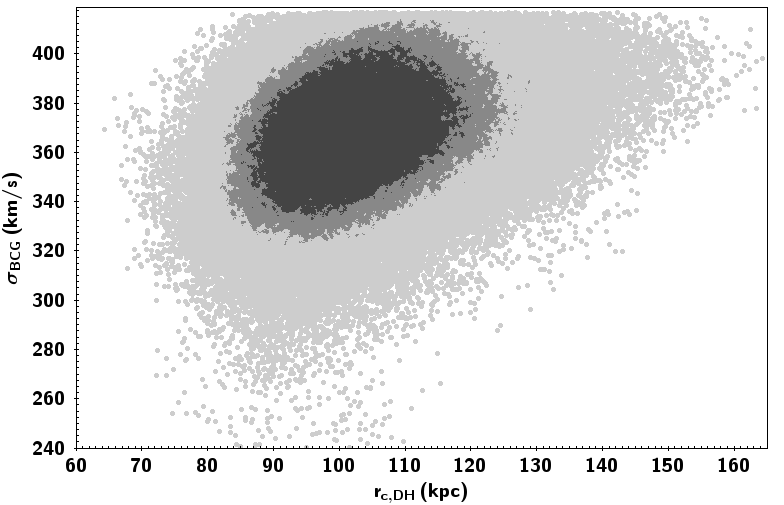}
 \caption{\small Large scale DH core radius versus the velocity dispersion of the dPIE mass profile describing the cluster BCG. The gray scale correspond to the $1\sigma$ (black), $2\sigma$ (dark gray) and $3\sigma$ (light gray) confidence level region.}  
         \label{fig:dh_bcg}
 \end{figure}

\begin{figure}
 \centering
 \includegraphics[width=8.5cm]{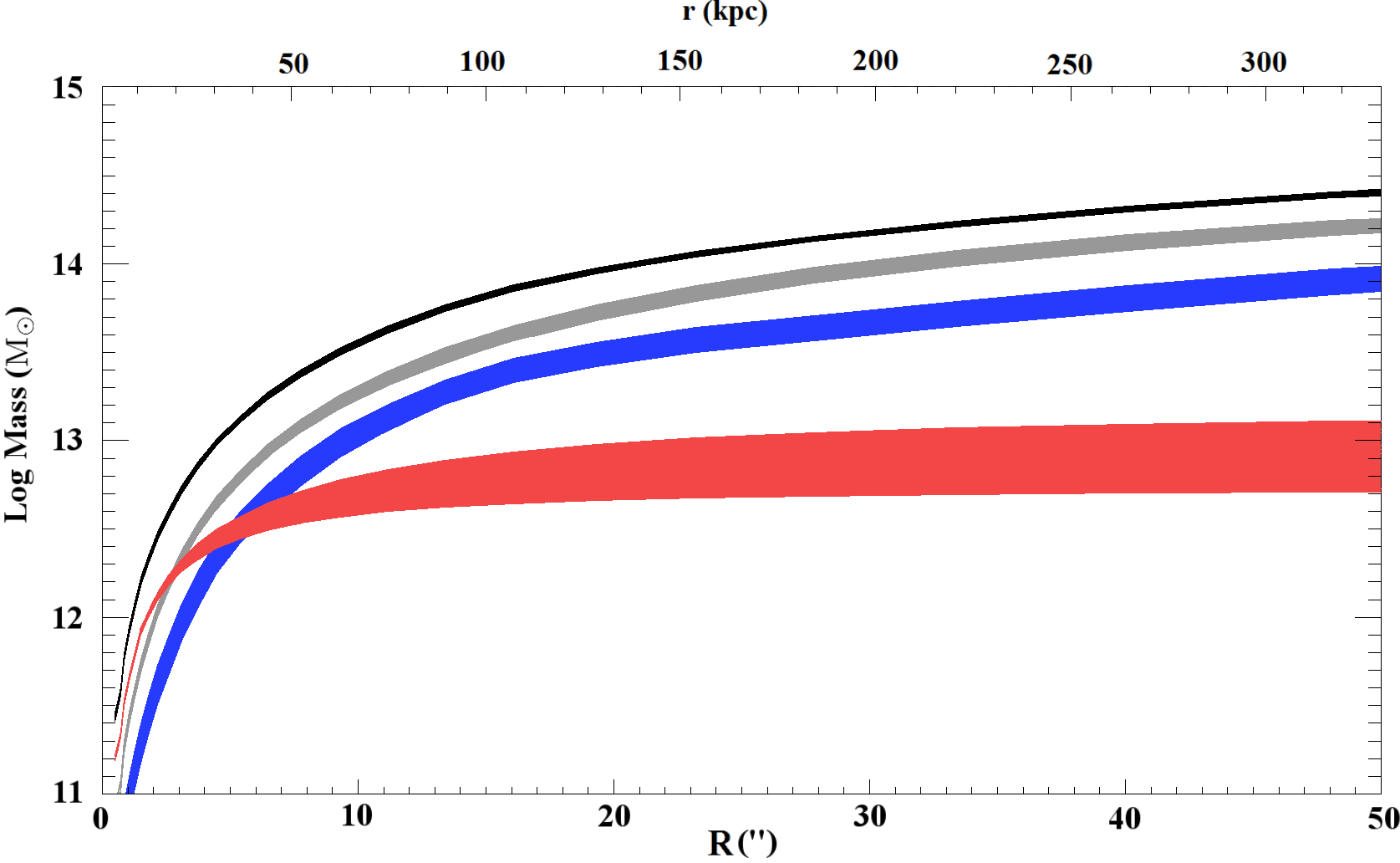}
 \caption{\small Projected 2D mass profile at 68\% confidence level of the core of MACS\,2129. In red we plot the BCG stellar mass profile, in blue we plot the total mass of the other cluster members, in gray the cluster DH, and in  black the total mass profile.}  
         \label{fig:mass_prof}
 \end{figure}

In Tab.\,\ref{tab:mass_param} we provide the final values obtained for the cluster mass components parameters.   \\
The cluster DH has central position very close to the BGC, with $\rm x_{DH}=-2.0\arcsec\pm0.7\arcsec$ and $\rm y_{DH}=1.5\arcsec\pm0.3\arcsec$ with respect to the BCG. It is highly elongated along the WE direction, with axis ratio $b/a=0.32\pm0.03$ and position angle PA$=-7^{\circ}\pm1^{\circ}$, measured counterclockwise from the west direction. The Einstein parameter is $\rm \Theta_e=29\pm4$ arcseconds, which corresponds to an Einstein radius of $\rm \theta_{E,sys1}=14\pm2$ arcseconds for a source at the redshift of system 1, $\rm z_{sys1}=1.36$. The DH has a quite large core radius of $\rm r_c= 101_{-11}^{+13}$ kpc.
Prediction of large DH core radii for galaxy clusters is not an unusual result in strong lensing analyses \citep[see e.g. ][]{ Richard2010a, Johnson2014,  Jauzac2014, Jauzac2015, Grillo2014}.  
In the innermost region of the cluster, lensing probes the total mass of the cluster core. This  is given by the superposition of the PIEMD profile describing the large scale DH and the dPIE profile which describes the baryonic mass component of the BCG. Obviously the core of the large scale DH profile correlates with the baryonic mass distribution of the BCG. The amplitude of the BCG stellar mass profile is described by  a \textit{central velocity dispersion $\rm\sigma_{0,BCG}^{dPIE}$} parameter, for the dPIE profile. This does not correspond to the spectroscopic measured central stellar velocity dispersion of the galaxy $\rm\sigma_{sp}^{*}$, since in this case $\rm\sigma_{0,BCG}^{dPIE}$ simply provide an estimate of the mass amplitude of the BCG stellar component, and is not related to the dynamical velocity dispersion of the galaxy. This can be seen in \citet{Newman2013a, Newman2013b} where, by combining lensing and kinematic analyses, they derive the amplitude $\rm\sigma_{0,BCG}^{dPIE}$ of the BCG stellar mass profile for several galaxy clusters, and the results indeed depart from the spectroscopic measurement of the central velocity dispersion of  the BCGs. E.g., for the BCG of Abell\,611 they get $\rm\sigma_{0,BCG}^{dPIE}=164\pm33\,km/s$ for the baryonic dPIE profile, whereas the spectroscopic measurement of the central velocity dispersion  provides $\rm\sigma_{BCG, sp}^{*}=317\pm20\,km/s$.
 \citealt{Sand2002, Sand2004} and \citealt{Newman2013a, Newman2013b} show the utility to combine the BCG stellar kinematic analysis with strong lensing to disentangle the DH and the stellar mass components in the innermost region of galaxy clusters.
\citealt{Newman2013b} separate the baryonic and dark matter components in seven galaxy clusters. They find that the size and mass of the BCG correlate with the core radii of the DH mass profile, and that larger BCG are hosted by clusters with larger core radii.  However they obtain typical DH core radii $\rm r_c\approx 10-20 kpc$ for the clusters of their sample. Thus the large core radius resulting in our analysis may reflect an overestimate of the mass associated to the BCG component.
Indeed from our analysis we get  $\rm\sigma_{0,BCG}^{dPIE}=370\pm25\,km/s$ as amplitude of the BCG stellar mass profile, which is rather high when compared to the results from  \citet{Newman2013a, Newman2013b} for the BCG stellar profiles. In order to proper disentangle the large scale DH and the BCG baryonic mass components in the core of the cluster, the lensing analysis alone is not enough, and additional constraints, as kinematics, are needed. In Fig. \ref{fig:dh_bcg} we plot the DH core radius versus the dPIE velocity dispersion of the BCG stellar mass profile for our mass model to show the correlation between this two parameters. In Fig.\,\ref{fig:mass_prof} we also show the mass profile for both this mass components. Given the total mass profile of the cluster robustly probed by the lensing analysis, we can see that a higher BCG stellar mass profile would imply a shallower profile for the DH, i.e. a larger DH core.
In addition we verified that, if we did not use a separate mass profile to describe the BCG and the inner most cluster mass would have been modeled by a PIEMD profile alone, then we would obtain a cluster core radius as small as $\rm \approx 5 kpc$, which  places a lower limit to the DH core. \\ 

Given our final best model of the cluster, we extract the projected mass of the different cluster components within the Einstein parameter. The projected mass of the large scale DH is constrained with 8\% precision, with  $\rm M_{DH}=(8.6\pm0.6)\times10^{13}\rm{M_\odot}$ within  $\rm \Theta_{E}=29\pm4$\arcsec. The BCG has a projected baryonic mass of $\rm M_{BCG}(<\Theta_E)=(8.4\pm2)\times10^{12}\rm{M_\odot}$ .\\
The reference galaxy GR has  central velocity dispersion $\sigma_{0,GR}=185_{-16}^{+	20}$ km/s and truncation radius $\rm r_{\rm tr, GR}=	66_{-	28	}^{+	22}$ kpc, providing the following cluster scaling relation:
\begin{equation}
\rm r_{\rm tr}=66_{-	28	}^{+	22} kpc\left(\frac{\sigma}{185_{-16}^{+	20}\,km/s}\right)^\frac{4}{3}\,.
\label{eq:scaling_rel}
\end{equation}
The other two galaxies individually optimized, G1 and G2, get velocity dispersion with $1\sigma$ errors of $\lesssim20\%$,  but their truncation radii have very large uncertainties.
The total projected mass of the other cluster galaxies is constrained at level of $12\%$, with $\rm M_{gal}=(4.4\pm0.5)\times10^{13}\rm{M_\odot}$ within the Einstein parameter $\rm \Theta_{E}$. \\
Overall the joint projected cluster mass of the large scale DH, BCG and the galaxy component  is $\rm M_{tot}( <\Theta_E)=(1.35\pm0.03)\times10^{14}\rm{M_\odot}$ and it is constrained with $2\%$ precision.
In Figure \ref{fig:mass_prof} we show the  2D projected mass profiles of the DH, BCG, the galaxy component and the total mass of the cluster core  up to a radial distance of $50\arcsec$ from the cluster center.  \\
\citet{Zitrin2015} published a first model of MACS\,2129 using the CLASH dataset. They obtain a total projected mass of $\rm M_e=(9.2\pm0.9)\times10^{13}M_{\odot}$ within the Einstein radius $\theta_{E}=19\arcsec\pm2\arcsec$ for a source at redshift $\rm z_s=2$. Within the same radius we extract a total mass of  $\rm M_{tot}(<19\arcsec)=(8.9\pm0.1)\times10^{13}M_{\odot}$, which is fully consistent with the previous analysis from \citet{Zitrin2015}. We highlight that the larger  uncertainties in \citet{Zitrin2015} account also for systematics, derived by modelling the DH cluster with different analytic mass profiles.

\begin{table}
\centering
\caption{\small Final parameters of the mass components of MACS\,2129 with their respective $1\sigma$ uncertain. 
The DH position (x,y) is given in arcseconds with respect to the BCG. The core and truncation radii are in kpc, velocity dispersions are in km/s and PA are in degrees measured counterclockwise from the west direction.}
\footnotesize
\begin{tabular}{|l|c|}
\hline
\hline
  DH & \\         															
 \hline            															
x	& $	-2.0\pm0.7$\\
y	& $	1.5	\pm0.3$\\
b/a	& $	0.32	\pm0.03$\\
PA	& $	-7\pm1$\\
 $\Theta_e (")$  	& $	29	\pm4$\\
 $\rm r_{c}$& $	101	_{-	11	}^{+	13	}$\\
 \hline  
GR	& \\
 \hline
$\sigma_0$& $185_{-16}^{+20}$\\
$\rm r_{\rm tr}$	& $	66	_{-	28	}^{+	22	}$\\
 \hline  
 BCG	&\\
 \hline  
 b/a	& $	0.78	\pm0.10$\\
PA	& $	2.5\pm2.8$\\
$\sigma_0$& $370\pm25$\\
 $\rm r_{c}$	& $	1.1	_{-	0.8}^{+	1.2	}$\\
$\rm r_{\rm tr}$	& $	87	_{-	35	}^{+	32	}$\\
 \hline  
G1	& \\
 \hline  
b/a	& $	0.86\pm0.05$\\
PA	& $	54\pm3$\\
$\sigma_0$& $224_{-38}^{+52}$\\ 
$\rm r_{c}$	& $1.6	_{-	1.0	}^{+	1.1	}$\\
$\rm r_{\rm tr}$	& $	57	_{-	37	}^{+	30	}$\\
 \hline  
G2	& \\
 \hline  
b/a	& $	0.80\pm0.05$\\
PA	& $	130\pm3$\\
$\sigma_0$& $255\pm28$\\
 $\rm r_{c}$	& $	1.3_{-	1.0}^{+	1.3	}$\\
$\rm r_{\rm tr}$	& $	25_{-	21	}^{+	28	}$\\
 
 \hline     
\hline     
  \end{tabular}\\
\label{tab:mass_param}
\end{table}

\section{Summary and Conclusions}
\label{sec:conclusions}

In this paper we presented a precise strong lensing analysis of the massive cluster of galaxies MACS\,2129. We combine the CLASH high resolution photometry with the new VLT spectroscopy from the  CLASH-VLT surveys to robustly select cluster members and multiple images needed for the SL modeling. This cluster shows a peculiar sextuply lensed red passive galaxy in its core, which has spectroscopic redshift $\rm z_{sp}=1.36$. 
Six additional multiple lensed sources are clearly identified in the cluster core. Our CLASH-VLT spectroscopic data provide spectroscopic redshift confirmation for three of these lensed systems.
We use these multiple images  to constrain the mass profile in the inner most region of the cluster with very high precision.  Our best mass reconstruction of the cluster reproduces the observed multiple images with a median accuracy of $0.4\arcsec$. 
The overall high accuracy in predicting the position of the multiple images  directly translate to a high precision reconstruction of the total mass distribution of the cluster.\\
The cluster DH projected mass is constrained with 8\% precision, with $\rm M_{DH}=(8.6\pm0.6)\times10^{13}\rm M_{\odot}$  within the Einstein parameter $\Theta_E=29\arcsec\pm4\arcsec$. The DH has a large core radius $\rm r_c=101_{-11}^{+13} kpc$, similar to other galaxy cluster \citep{Richard2010, Johnson2014, Jauzac2014, Jauzac2015, Grillo2014}. However, the DH core radius correlates with the BCG mass profile. When the BCG baryonic and the cluster  DH components are properly disentangled by combining lensing and kinematic analyses, typical smaller core radii $(\rm r_c\approx 10-20 kpc)$ are found for cluster DHs, see \citealt{Newman2013a, Newman2013b}. Thus the large core radius we derived likely reflects an overestimate of the mass associated to the BCG mass component.  Nevertheless the final 2D projected total mass of the cluster is constrained with an accuracy of 2\%. Within the Einstein parameter it is $\rm M_{tot}=(1.35\pm0.03)\times10^{14}\rm M_{\odot}$. \\
 In addition to the mass model for the core of MACS\,2129, we also provide the complete list of robust spectroscopic redshifts of cluster members and lensed features measured with VLT/VIMOS within the CLASH-VLT survey, see  the Appendix \ref{sec:appendix}.
\section*{Acknowledgements}

This work is supported by the Transregional Collaborative Research Centre TRR 33 - The
Dark Universe and the DFG cluster of excellence ``Origin and Structure of the Universe". 
The CLASH Multi-Cycle Treasury Program (GO-12065) is based on observations made with the NASA/ESA Hubble Space Telescope. The Space Telescope Science Institute is operated by the Association of Universities for Research in Astronomy, Inc. under NASA contract NAS 5-26555. 
SHS gratefully acknowledges support from the Max Planck Society through the Max Planck Research Group.
P.R. and A. Mercurio acknowledge support
from PRIN-INAF 2014 1.05.01.94.02 (PI M. Nonino). 
P.R. acknowledges the hospitality and support of the visitor program of the DFG cluster of excellence  ``Origin and Structure of the Univers''. 
C.G. acknowledges support by VILLUM FONDEN Young Investigator Programme through grant no. 10123.

\addcontentsline{toc}{chapter}{Bibliography}
\bibliographystyle{mn2efix}
\bibliography{macs2129}

\appendix
\section{Cluster members and lensed images spectroscopy}
\label{sec:appendix}
In this section we provide the spectroscopic information from the CLASH-VLT survey for all the confirmed cluster members (in Tab.~\ref{tab:galaxy_vlt}) as well as the list of all the lensed sources targeted with VLT, including the multiple images systems 1 to 4 and system 8, used in the strong lensing analysis (see Tab.~\ref{tab:lensed_images}).\\
We remark that only two slits masks were used with the VIMOS MR grism due to strong galactic cirrus in MACS\,2129, resulting in a small number of redshifts compared to other CLASH-VLT clusters.

\begin{table}
\caption{List of the spectroscopically confirmed  lensing features in the core of MACS\,2129: column (1) id, columns (2) and (3) RA and DEC in degrees, column (4) spectroscopic redshift, column (5)  quality flag. The first six objects are multiple images used in the strong lensing model.}
\centering
\footnotesize
\begin{tabular}{|c|c|c|c|c|}
\hline
\hline
ID& Ra & DEC& $\rm z_{sp}$& QF \\
\hline
 1.1 &  322.358001  &-7.685835  &1.3649  &2 \\
 1.5 &  322.357147  &-7.691137  &1.3641  &3 \\
 2.1 &  322.354838  &-7.69073   &1.0400  &2 \\
 3.1 &  322.35027   &-7.688832  &2.2362  &2 \\
 3.3 &  322.350971  &-7.695714  &2.2392  &2 \\
 4.1 &  322.366451  &-7.686847  &2.2367  &2 \\
 8 &  322.35698   &-7.689239  &4.4107  &9 \\
9  &    322.36302   &-7.685172  &1.4519  &9 \\
10 &     322.368227 & -7.692496 & 1.5453 & 3\\
11 &     322.361659 & -7.696394 & 0.9484 & 3\\
12 &     322.354701 & -7.700377 & 2.4129 & 2\\
13 &     322.348531 & -7.699152 & 0.9474 & 3\\
14 &322.362804	&-7.683168	&2.3082	&2 \\
15 &322.34094	&-7.684378	&2.2707	&3 \\
16 &322.360655	&-7.70458	&1.9939	&2 \\
17 &322.354359	&-7.705515	&1.1815	&3 \\
18 &322.355495	&-7.707085	&1.0511	&3 \\
19 &322.370793	&-7.709214	&1.0495	&3 \\
20 &    322.350806   &-7.681908  &1.3599  &3 \\
21 &    322.36481    &-7.683756  &1.3698  &3 \\
22  &     322.362883 & -7.68268  & 2.3082 & 2\\
\hline                  
\end{tabular}
\label{tab:lensed_images}
\end{table}

\begin{table}
\caption{List of MACS\,2129 cluster members spectroscopically confirmed by VLT/VIMOS data: column (1) id, columns (2) and (3) RA and DEC in degrees, column (4) VIMOS spectroscopic redshift, column (5)  Quality Flag.}
\centering
\footnotesize
\begin{tabular}{|c|c|c|c|c|}
\hline
\hline
ID& Ra & DEC& $\rm z_{sp}$& QF \\
\hline
  10481   &  322.189736 & -7.65697  & 0.5887   &3    \\
  10794   &  322.347084 & -7.663556 & 0.5922   &3    \\
  10916   &  322.36514  & -7.666103 & 0.5937   &3    \\
  10930   &  322.353041 & -7.666496 & 0.5942   &3    \\
  11120   &  322.327031 & -7.67048  & 0.5927   &3    \\
  11141   &  322.337188 & -7.670573 & 0.5959   &3    \\
  11422   &  322.311935 & -7.676886 & 0.5865   &3    \\
  11760   &  322.340785 & -7.683884 & 0.5817   &2    \\
  12388   &  322.357097 & -7.689107 & 0.5833   &3    \\
  12431   &  322.201623 & -7.69605  & 0.5948   &3    \\
  12475   &  322.350967 & -7.697143 & 0.5933   &3    \\
  12489   &  322.245821 & -7.697805 & 0.5835   &3    \\
  12490   &  322.245601 & -7.698231 & 0.5807   &3    \\
  13051   &  322.353276 & -7.707495 & 0.5746   &3    \\
  13052   &  322.353245 & -7.707943 & 0.5904   &3    \\
  13061   &  322.290916 & -7.707586 & 0.5961   &3    \\
  13217   &  322.333338 & -7.710552 & 0.5928   &3    \\
  13373   &  322.325793 & -7.714155 & 0.5832   &3    \\
  13652   &  322.360905 & -7.719659 & 0.5886   &2    \\
  13694   &  322.316236 & -7.720489 & 0.578    &3    \\
  13886   &  322.350413 & -7.72435  & 0.5738   &2    \\
  14049   &  322.394194 & -7.728681 & 0.5779   &3    \\
  14147   &  322.402022 & -7.729684 & 0.5999   &3    \\
  14698   &  322.303043 & -7.741079 & 0.5906   &3    \\
  14851   &  322.145996 & -7.744111 & 0.5868   &3    \\
  16408   &  322.34314  & -7.782107 & 0.5901   &3    \\
  17053   &  322.211216 & -7.795855 & 0.5894   &2    \\
  17545   &  322.321102 & -7.806902 & 0.5788   &3    \\
  17758   &  322.192709 & -7.811686 & 0.5934   &3    \\
  4248    &  322.165238 & -7.892864 & 0.5851   &3    \\
  4351    &  322.311384 & -7.890688 & 0.5854   &3    \\
  4506    &  322.151681 & -7.88787  & 0.5952   &3    \\
  4518    &  322.151734 & -7.886972 & 0.5955   &3    \\
  4588    &  322.139294 & -7.886331 & 0.5814   &3    \\
  5525    &  322.169704 & -7.863769 & 0.5737   &3    \\
  5652    &  322.304068 & -7.860543 & 0.5869   &3    \\
  6351    &  322.227851 & -7.845947 & 0.5764   &3    \\
  6485    &  322.298203 & -7.843324 & 0.5809   &3    \\
  6529    &  322.210264 & -7.842364 & 0.5858   &3    \\
  6558    &  322.362822 & -7.842443 & 0.5777   &3    \\
  6830    &  322.322366 & -7.835512 & 0.5875   &3    \\
  7255    &  322.22226  & -7.824324 & 0.5744   &2    \\
  7359    &  322.189632 & -7.822517 & 0.5947   &3    \\
  9492    &  322.228274 & -7.634417 & 0.5834   &3    \\
  9744    &  322.342228 & -7.639917 & 0.5841   &3    \\
  13198   &  322.20745  & -7.710083 & 0.5843   &2    \\
  5448    &  322.20839  & -7.861958 & 0.58785  &3    \\
  12133   &  322.354958 & -7.691768 & 0.5959   &3    \\
\hline                  
\end{tabular}
\label{tab:galaxy_vlt}
\end{table}

\newpage
 \begin{figure*}
 \centering
 \includegraphics[width=18cm]{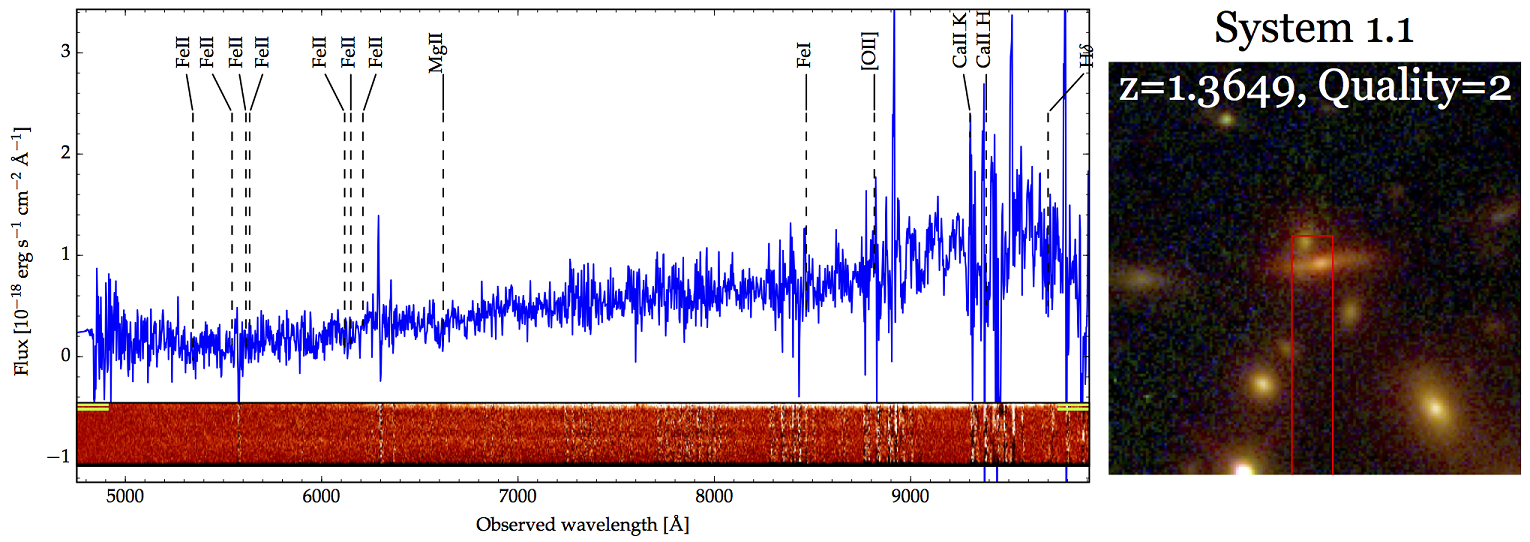}
  \includegraphics[width=18cm]{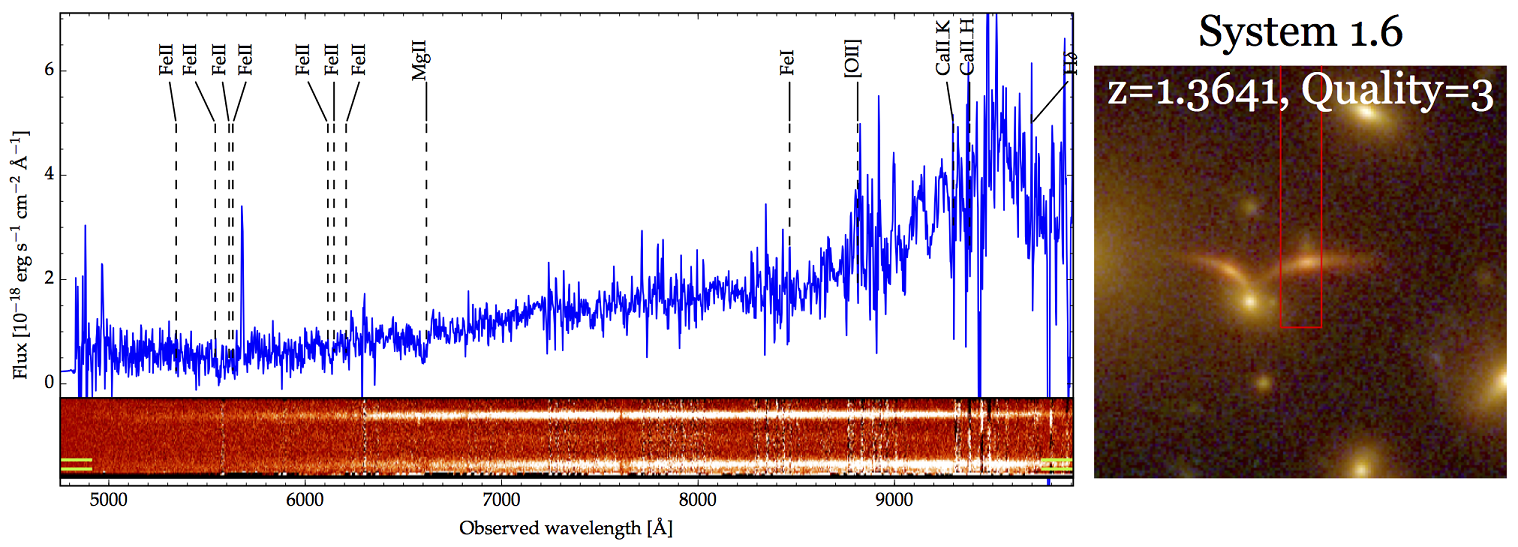}
   \includegraphics[width=18cm]{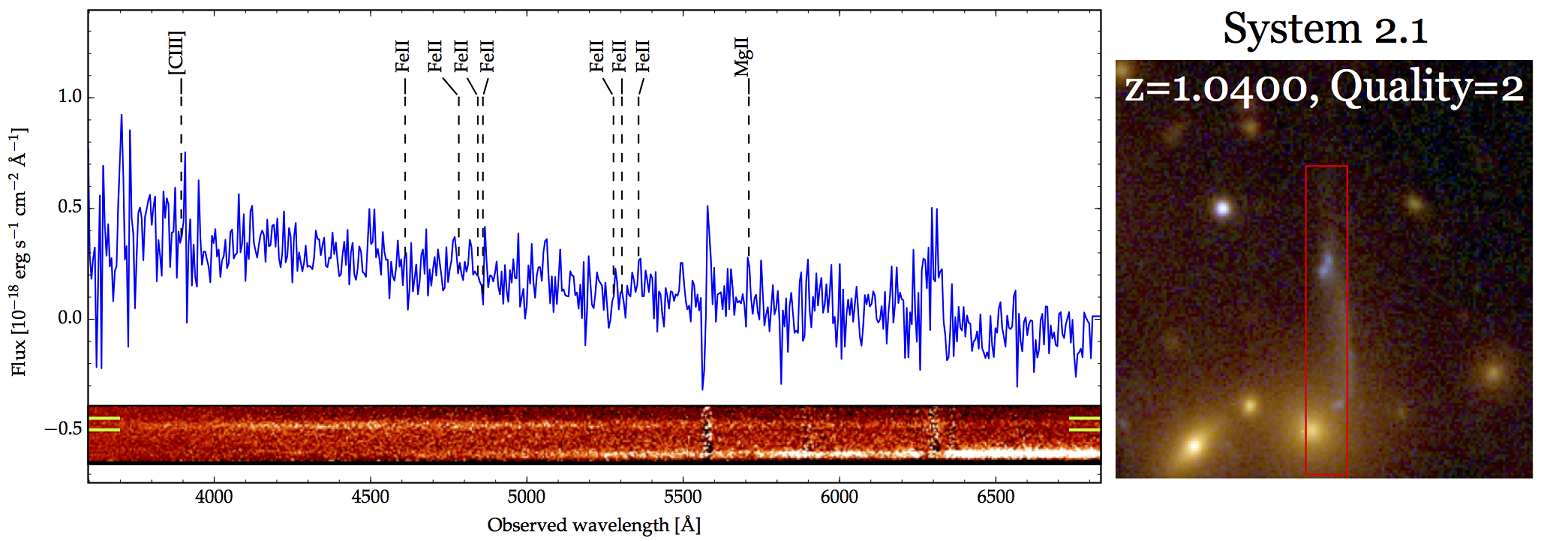}
 \caption{\small Spectroscopy of the multiple images of system 1 and 2 observed with VLT/VIMOS. On the left we present the 2D (lower panel)  and 1D spectra (upper panel). The dashed vertical lines in the 1D spectra mark emission and absorption lines used to derive the source redshift. On the right we show the HST color composite image cutout of the observed multiple image. The red rectangles mark the position of the 1\arcsec-long VIMOS slits.  In the 2D spectra we  use two green lines to indicate the position of the emission associated to the lensed image.  }  
         \label{fig:spectra1}
 \end{figure*}
 \begin{figure*}
 \centering
 \includegraphics[width=18cm]{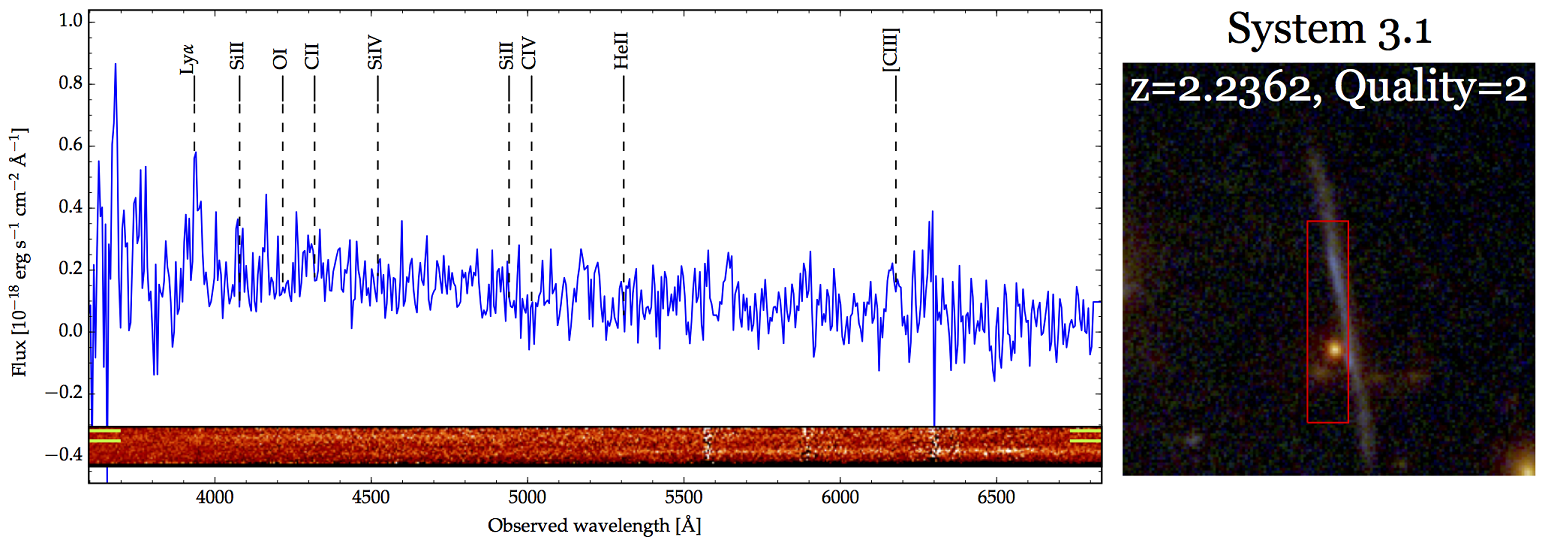}
  \includegraphics[width=18cm]{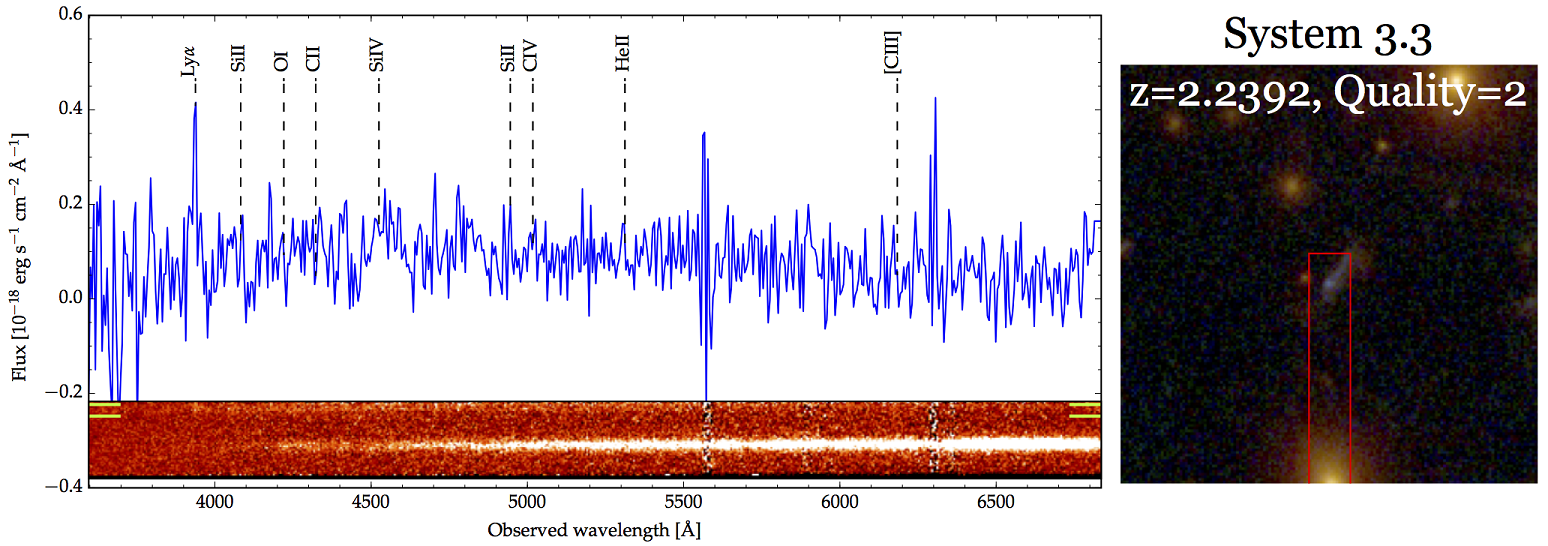}
   \includegraphics[width=18cm]{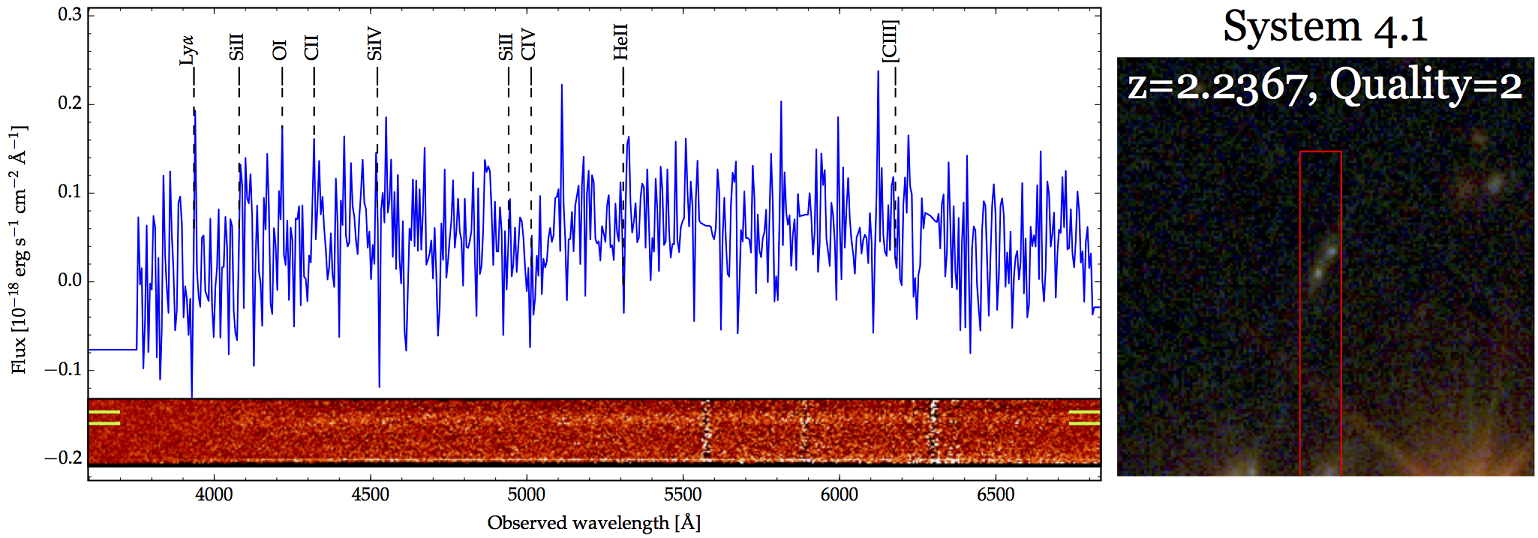}
 \caption{\small Same as Fig.\,\ref{fig:spectra1}, for the multiple images systems 3 and 4.}  
         \label{fig:spectra2}
 \end{figure*}
\begin{figure*}
 \centering
 \includegraphics[width=18cm]{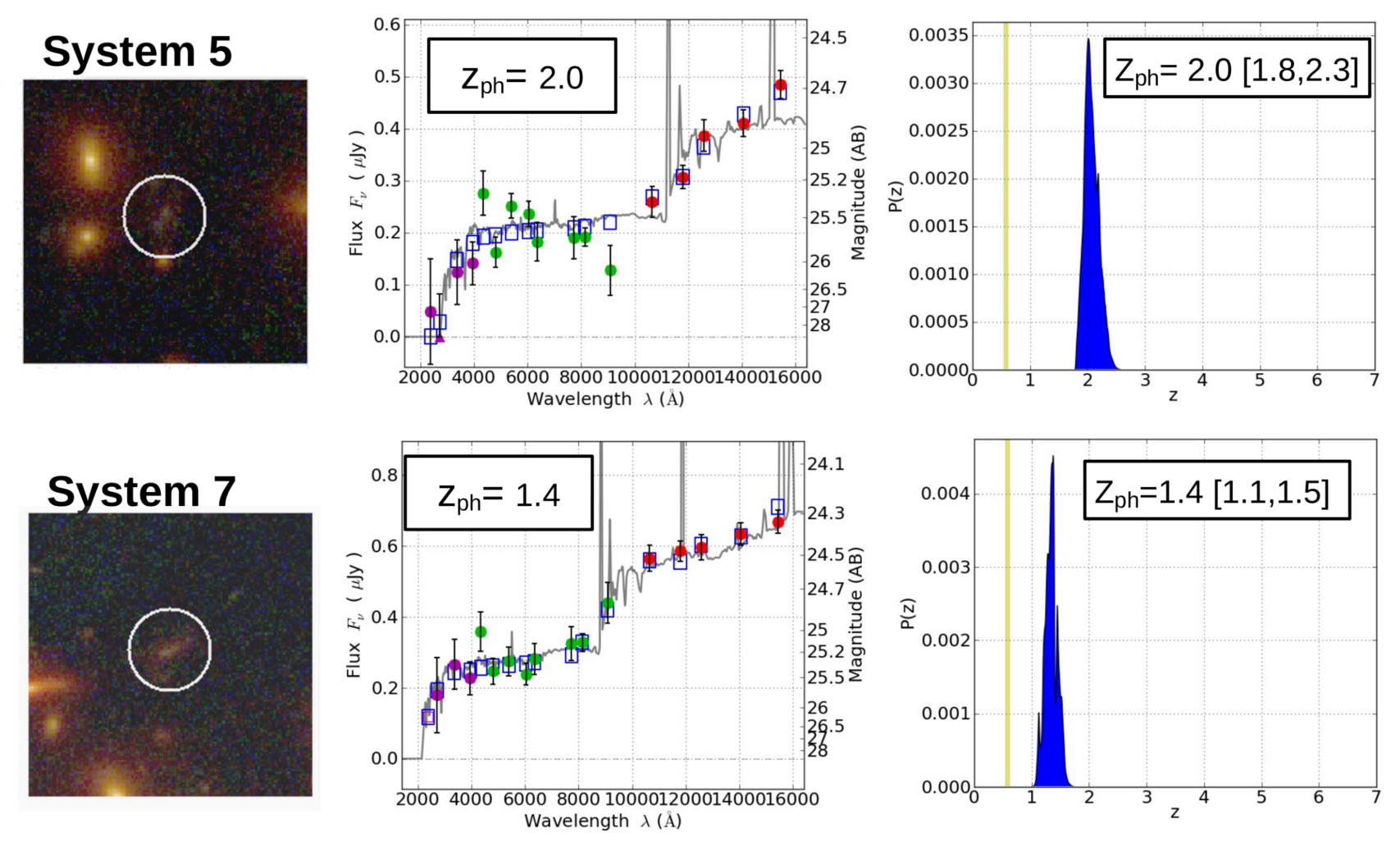}
 \caption{\small Spectral Energy Distribution (SED) for the multiple images systems 5 and 7, derived using the CLASH photometry and the SED fit code BPZ. We present the fit of the multiple images 5.2  and 7.3, which get well defined PDF(z). On the left we show the lensed images cutout in the color composite image of the cluster. The white circle label the image and has diameter of $\sim1.5$ arcseconds. The central panel shows the SED fit with the photometric measurement from CLASH, showing in violet the fluxes in the WFC3UVIS filter, in green the one in the ACS , and in red the IR photometry observed with the WFC3IR. The gray line shows the best template fitting the photometry, and in the legend we provide the respective best fitting redshift. On the right we show the respective PDF(Z) and provide the best fitting $z_{ph}$ with the respective 68\% confidence level interval. The vertical yellow line marks the cluster redshift. }  
         \label{fig:dan_phot}
 \end{figure*}
 \begin{figure*}
 \centering
 \includegraphics[width=18cm]{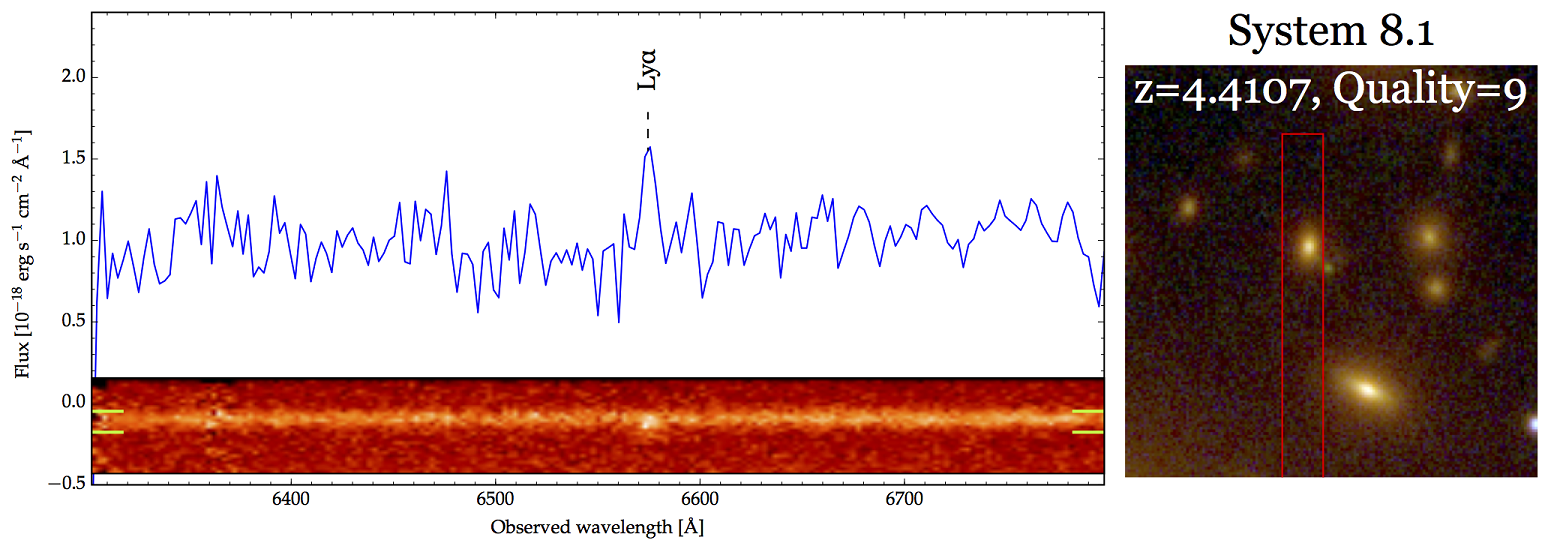}
 \caption{\small Same as Fig.\,\ref{fig:spectra1} for the multiple lensed image 8.1 at redshift $\rm z_{sp}=4.4$. The lensed image is the green bright source lying close to the slit edge in the RGB cutout.}  
         \label{fig:spectra_z4}
 \end{figure*}

\bsp

\label{lastpage}

\end{document}